\documentstyle[psfig]{mn}

\newif\ifAMStwofonts

\newcommand{\beq}{\begin{equation}}
\newcommand{\eeq}{\end{equation}}
\newcommand{\beqn}{\begin{eqnarray}}
\newcommand{\eeqn}{\end{eqnarray}}
\def\bI{\hbox{$\,I\!\!\!\!-$}}
\def\agt{\mathrel{\raise.3ex\hbox{$>$}\mkern-14mu\lower0.6ex\hbox{$\sim$}}}
\def\alt{\mathrel{\raise.3ex\hbox{$<$}\mkern-14mu\lower0.6ex\hbox{$\sim$}}}

\ifoldfss
  \ifCUPmtlplainloaded \else
    \NewTextAlphabet{textbfit} {cmbxti10} {}
    \NewTextAlphabet{textbfss} {cmssbx10} {}
    \NewMathAlphabet{mathbfit} {cmbxti10} {} 
    \NewMathAlphabet{mathbfss} {cmssbx10} {} 
  \fi
  \ifAMStwofonts
    \ifCUPmtlplainloaded \else
      \NewSymbolFont{upmath} {eurm10}
      \NewSymbolFont{AMSa} {msam10}
      \NewMathSymbol{\upi}     {0}{upmath}{19}
      \NewMathSymbol{\umu}     {0}{upmath}{16}
      \NewMathSymbol{\upartial}{0}{upmath}{40}
      \NewMathSymbol{\leqslant}{3}{AMSa}{36}
      \NewMathSymbol{\geqslant}{3}{AMSa}{3E}

      \let\leq=\leqslant 
      \let\geq=\geqslant 
    \fi
  \fi
\fi 

\ifnfssone
  \newmathalphabet{\mathit}
  \addtoversion{normal}{\mathit}{cmr}{m}{it}
  \addtoversion{bold}{\mathit}{cmr}{bx}{it}
  \newmathalphabet{\mathbfit} 
  \addtoversion{normal}{\mathbfit}{cmr}{bx}{it}
  \addtoversion{bold}{\mathbfit}{cmr}{bx}{it}
  \newmathalphabet{\mathbfss} 
  \addtoversion{normal}{\mathbfss}{cmss}{bx}{n}
  \addtoversion{bold}{\mathbfss}{cmss}{bx}{n}
  \ifAMStwofonts
    \ifCUPmtlplainloaded \else
      \UseAMStwoboldmath
      \makeatletter
      \new@mathgroup\upmath@group
      \define@mathgroup\mv@normal\upmath@group{eur}{m}{n}
      \define@mathgroup\mv@bold\upmath@group{eur}{b}{n}
      \edef\UPM{\hexnumber\upmath@group}
      \new@mathgroup\amsa@group
      \define@mathgroup\mv@normal\amsa@group{msa}{m}{n}
      \define@mathgroup\mv@bold\amsa@group{msa}{m}{n}
      \edef\AMSa{\hexnumber\amsa@group}
      \makeatother
      \mathchardef\upi="0\UPM19
      \mathchardef\umu="0\UPM16
      \mathchardef\upartial="0\UPM40
      \mathchardef\leqslant="3\AMSa36
      \mathchardef\geqslant="3\AMSa3E

      \let\leq=\leqslant 
      \let\geq=\geqslant 
    \fi
  \fi
\fi 

\ifnfsstwo
  \DeclareMathAlphabet{\mathbfit}{OT1}{cmr}{bx}{it}
  \SetMathAlphabet\mathbfit{bold}{OT1}{cmr}{bx}{it}
  \DeclareMathAlphabet{\mathbfss}{OT1}{cmss}{bx}{n}
  \SetMathAlphabet\mathbfss{bold}{OT1}{cmss}{bx}{n}
  \ifAMStwofonts
    \ifCUPmtlplainloaded \else
      \DeclareSymbolFont{UPM}{U}{eur}{m}{n}
      \SetSymbolFont{UPM}{bold}{U}{eur}{b}{n}
      \DeclareSymbolFont{AMSa}{U}{msa}{m}{n}
      \DeclareMathSymbol{\upi}{0}{UPM}{"19}
      \DeclareMathSymbol{\umu}{0}{UPM}{"16}
      \DeclareMathSymbol{\upartial}{0}{UPM}{"40}
      \DeclareMathSymbol{\leqslant}{3}{AMSa}{"36}
      \DeclareMathSymbol{\geqslant}{3}{AMSa}{"3E}

      \let\leq=\leqslant 
      \let\geq=\geqslant 
    \fi
  \fi
\fi 

\ifCUPmtlplainloaded \else
  \ifAMStwofonts \else 
    \def\upi{\pi}
    \def\umu{\mu}
    \def\upartial{\partial}
  \fi
\fi

\title{Dynamical instability of differentially rotating stars}
\author[Shibata et al.]
       {Masaru Shibata, Shigeyuki Karino, and Yoshiharu Eriguchi \\ 
	Department of Earth Science and Astronomy, 	
	Graduate School of Arts and Sciences,~University of Tokyo,\\
	Komaba, Meguro, Tokyo 153-8902, Japan}
\date{Accepted ???? Month ??,
      Received ???? Month ??;
      in original form 2002 April 1}

\pagerange{\pageref{firstpage}--\pageref{lastpage}}
\pubyear{????}

\begin{document}

\maketitle

\label{firstpage}

\begin{abstract}
We study the dynamical instability against bar-mode
deformation of differentially rotating stars. 
We performed numerical simulation and linear perturbation analysis
adopting polytropic equations of state with the
polytropic index $n=1$. It is found that 
rotating stars of a high degree of 
differential rotation are dynamically unstable
even for the ratio of the kinetic energy to the gravitational
potential energy of $O(0.01)$. Gravitational waves 
from the final nonaxisymmetric quasistationary states are calculated 
in the quadrupole formula. For rotating stars of mass $1.4M_{\odot}$ and
radius several 10 km, gravitational waves have frequency 
several $100$ Hz and effective amplitude $\sim 5 \times 10^{-22}$ 
at a distance of $\sim 100$ Mpc. 
\end{abstract}

\begin{keywords}
gravitational waves -- stars: neutron -- stars: rotation
-- stars: oscillation.
\end{keywords}

\section{INTRODUCTION}
Stars in nature are rotating and subject to nonaxisymmetric 
rotational instabilities.  An exact treatment of these instabilities 
exists only for incompressible equilibrium fluids in Newtonian 
gravity \cite{CH69,TA78}. For these configurations, global 
rotational instabilities arise from nonradial toroidal modes 
$e^{im\varphi}$ ($m=\pm 1,\pm 2, \dots$) when $\beta\equiv T/W$ exceeds a 
certain critical value. Here $\varphi$ is the azimuthal coordinate and 
$T$ and $W$ are the rotational kinetic and gravitational potential 
energies.  In the following we will focus on the $m=\pm 2$ bar-mode 
since it is expected to be the fastest growing mode
(but see Centrella et al. (2001) with regard to a counter-example 
for extremely soft equations of state). 

There exist two different mechanisms and corresponding timescales for 
bar-mode instabilities.  Uniformly rotating, incompressible stars in 
Newtonian theory are {\em secularly} unstable to bar-mode deformation 
when $\beta \geq \beta_s \simeq 0.14$.  However, this instability can 
only grow in the presence of some dissipative mechanism, like 
viscosity or gravitational radiation, and the growth time is 
determined by the dissipative timescale, which is usually much longer 
than the dynamical timescale of the system.  By contrast, a {\em 
dynamical} instability to bar-mode deformation sets in when $\beta \geq 
\beta_d \simeq 0.27$.  This instability is independent of any 
dissipative mechanisms, and the growth time is determined 
typically by the hydrodynamical timescale of the system. 

For the compressible case, 
determining the onset of the dynamical bar-mode instability, as well
as the subsequent evolution of an unstable star, requires numerical 
computations. Hydrodynamical simulations performed in Newtonian theory 
\cite{TDM,DGTB,WT,TH,CT2,CT3,CT1,PDD,Toman1,Toman2,New,brown,LL1,LL2} 
have shown that $\beta_d$ is $\sim 0.27$ 
as long as the rotational profile is not strongly differential. 
In this case, once a bar has 
developed, the formation of spiral arms plays an important role in 
redistributing the angular momentum and forming a core-halo structure.
Recently, it has been shown that 
$\beta_d$ can be smaller for stars with a higher degree 
of differential rotation \cite{CNLB,TH,PDD,LL1,LL2} as $\beta_d \sim 0.14$.
In such case, the formation of bars and spiral arms does not 
take place. Instead, small density perturbation is excited and 
left to be weakly nonlinear. To date, there is 
no report of dynamically unstable stars with $\beta \alt 0.14$.

There are numerous evolutionary paths which may lead to the formation 
of rapidly and differentially rotating neutron stars. 
$\beta$ increases approximately as $R^{-1}$ during stellar collapse. 
Also, with decreasing stellar radius, the 
angular velocity in the central region 
may be much larger than that in the outer region. 
During supernova collapse, the core contracts from $\sim 1,000$ km to 
several 10 km, and hence $\beta$ increases by two orders of magnitude. 
Thus, even rigidly rotating progenitor stars 
with a moderate angular velocity 
may yield rapidly and differentially rotating neutron stars 
which may reach the onset of dynamical instability \cite{BM,RMR}. 
Similar arguments hold for accretion induced collapse 
of white dwarfs to neutron stars and for the merger of binary white
dwarfs to neutron stars. 
Differential rotation is eventually suppressed by
viscous angular momentum transfer or magnetic braking \cite{BSS}, but
within the transport timescale, the differentially rotating 
stars are subject to nonaxisymmetric instabilities. 

In this Letter, we report some of results we have recently obtained on 
dynamical bar-mode instabilities in differentially rotating stars. 
We pay attention to rotating stars with a high degree of
differential rotation in Newtonian gravity 
using both linear perturbation analysis and 
nonlinear hydrodynamical simulation. 
The detail of our study will be summarized in
a future paper \cite{SKE}. Here, we highlight a new finding. 

\section{METHOD}

To investigate the dynamical stability, we first prepare rotating stars
in equilibrium. Rotating stars in equilibrium 
are modeled by the polytropic equations of state as 
$P=K\rho^{\Gamma}$ where $P$, $\rho$, $K$ and $\Gamma$ denote 
the pressure, density, polytropic constant and adiabatic index. 
Here, we choose $\Gamma=2$ to model moderately stiff equations of 
state for neutron stars. For simplicity, the 
profile of the angular velocity $\Omega$ is set as 
\beq
\Omega = {\Omega_0 A^2 \over \varpi^2 + A^2},
\eeq
where $A$ is a constant, $\Omega_0$ the angular velocity at 
the symmetric axis, and $\varpi$ cylindrical radius ($\sqrt{x^2+y^2}$).
$A$ controls the steepness of the rotational profile: For smaller
$A$, the profile is steeper and for $A \rightarrow \infty$,
the rigid rotation is recovered. 
In the limit of $A$ going to 0, this rotation law gives uniform specific 
angular momentum, which is the limiting law allowed by axisymmetric 
stability analyses. (Here, however, $A=0$ is prohibited in this situation
because angular velocity diverges along the rotation axis.)
This is also the rotation law that has been often used 
in studies of nonaxisymmetric instabilities in tori and annuli 
\cite{PP,GN,AT}.
In this Letter, 
we report the results for $\hat A \equiv A/R=0.3$ and 1
where $R$ is the equatorial radius of rotating stars.

In a linear perturbation analysis, we derive the
linear perturbation equations of the hydrodynamic equations and
substitute the functions in the form
$f(r, \theta)e^{im\varphi-i\omega t}$ for perturbed hydrodynamic
variables. As a result, the problem reduces to the eigen value
problem for determining a complex value of $\omega$ and
the corresponding eigen functions of the perturbed quantities.
The details of the numerical method will be reported in future \cite{KE}. 

In the hydrodynamical simulation, we 
initially superimpose a density perturbation 
of the bar-mode in the form 
\beq
\delta \rho = \delta \cdot \rho_0 {x^2 - y^2 \over R^2},
\eeq
where $\rho_0$ denotes the axisymmetric configuration and
$\delta$ constant. Namely, we focus on a fundamental mode
(f-mode) for which 
there is no node. In this work, we choose $\delta=0.1$, and 
the velocity is left to be unperturbed at $t=0$. 
The growth of a bar-mode can be followed by monitoring the distortion
parameter 
\beq
\eta \equiv {I_{xx} - I_{yy} \over I_{xx} + I_{yy}},
\eeq
where $I_{ij}$ denotes the quadrupole moment
\beq
I_{ij} = \int d^3x \rho x^i x^j. 
\eeq
Simulations were performed using a 3D numerical code
in Newtonian gravity \cite{SON}. We adopt 
a fixed uniform grid with 
size $141\times 141\times 141$ in $x-y-z$, which covers
an equatorial radius by 50 grid points initially. 
To confirm that the results depend weakly on the resolution,
we also perform test simulations with size 
$71\times 71\times 71$ (i.e., the grid spacing becomes twice larger)
for several selected cases. 
We assume a reflection symmetry about the equatorial plane. 
Since many rotating stars we pick up have flattened configuration, we 
set the grid spacing of $z$ half of that of $x$ and $y$. 

\section{NUMERICAL RESULTS}

First we show results for rotating stars of 
moderately large differential rotation with $\hat A=1$. 
In Fig. 1, we show $|\eta|$ as a function of time for 
$C_a=R_p/R=0.255$ and 0.305, where $R_p$ is the 
polar radius. It is found that for $C_a=0.255$, the rotating stars 
are dynamically unstable, while for $C_a=0.305$, they are stable. 
In Fig. 2, we also display $\omega_r/\Omega_0$ and $\omega_i/\Omega_0$
as a function of $\beta$. Here, $\omega_r$ was derived by 
carrying out the Fourier transformation of $\eta(t)$, and 
$\omega_i$ was measured using the growth rate of the peaks of $|\eta|$ 
(see dotted line in Fig. 1). 
We also compute these angular frequencies by a linear
perturbation analysis (solid lines). 
We find that two results independently obtained agree fairly well.
The linear analysis also indicates that this unstable mode 
is likely to be the f-mode (of no node in the eigen function). 

We find that at $C_a \sim 0.3$, the dynamical stability changes. 
In terms of $\beta$, 
the stability changes at $\beta\sim 0.26$ for $\hat A=1$. 
Thus, $\beta_d$ is not extremely smaller than that for rigidly rotating 
incompressible stars \cite{CH69} in the case when the
degree of the differential rotation is not very high.

\begin{figure}
\vspace*{-12mm}
\begin{center}
\leavevmode
\psfig{file=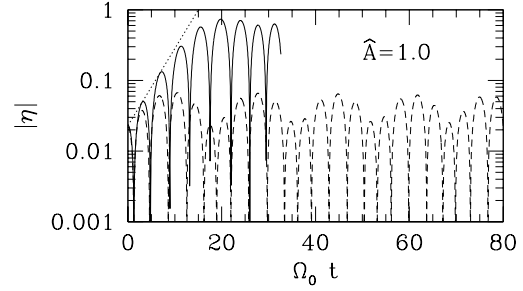,width=2.8in,angle=0}	
\end{center}
\vspace*{-22mm}
\caption{$|\eta|$ as a function of $\Omega_0 t$ for 
$\hat A=1$ with $C_a=0.255$ (solid line) and 0.305 (dashed line). 
The dotted line denotes the growth rate of the unstable mode. 
}
\end{figure}

\begin{figure}
\vspace*{-6mm}
\begin{center}
\leavevmode
\psfig{file=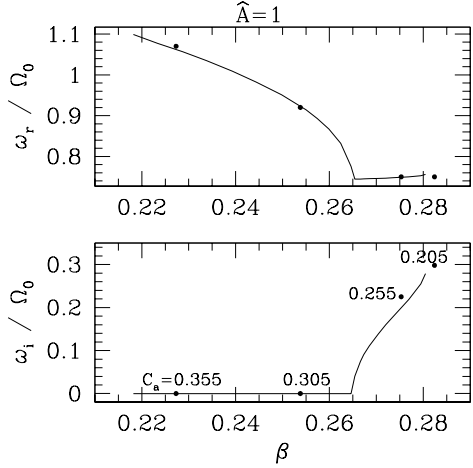,width=2.8in,angle=0}
\end{center}
\vspace*{-8mm}
\caption{$\omega_r/\Omega_0$ and $\omega_i/\Omega_0$
as a function of $\beta$ for $\hat A=1$.
The solid curves and solid circles denote
the results by linear perturbation analysis and numerical
simulation, respectively. We choose results for
$C_a=0.205$, 0.255, 0.305 and 0.355.  
}
\end{figure}

The situation drastically changes for $\hat A=0.3$. 
In Fig. 3, we show $|\eta|$ as a function of time for 
$C_a=0.405$, 0.605, and 0.805 obtained by 
numerical simulations. Note that for $C_a=0.805$,
$\beta \approx 0.04$. We find that in every model, 
nonaxisymmetric dynamical instabilities take place. 
In Fig. 4, we also show $\omega_r/\Omega_0$ and $\omega_i/\Omega_0$ 
as a function of $\beta$. Since several growing modes seem to be
excited simultaneously for $C_a=0.605$ and 0.805,
we plot upper and lower bounds for $\omega_i$.
Results by numerical simulations agree again with 
those by linear perturbation analysis. 
The linear analysis also indicates that even for $\beta \approx 0.03$,
the rotating stars are dynamically unstable. 

\begin{figure}
\vspace*{-4mm}
\begin{center}
\leavevmode
\psfig{file=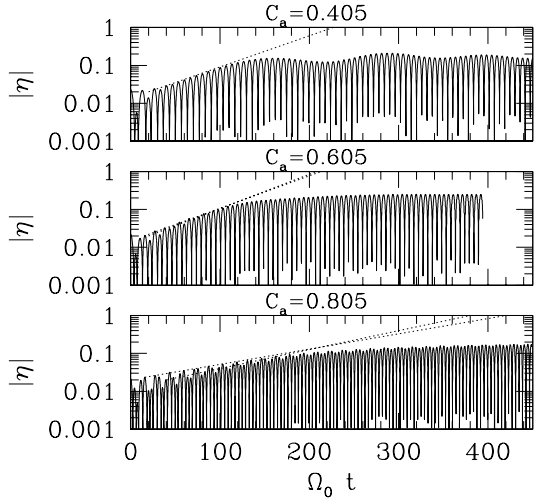,width=3.0in,angle=0}
\end{center}
\vspace*{-8mm}
\caption{$|\eta|$ as a function of $\Omega_0 t$ for 
$\hat A=0.3$ with $C_a=0.405$ (upper) 0.605 (middle), and 0.805 (lower). 
The dotted lines denote the growth rate of the unstable modes. 
For $C_a=0.605$ and 0.805, we plot two dotted lines which indicate
the upper and lower bounds of the growth rates. 
}
\end{figure}

\begin{figure}
\vspace*{-8mm}
\begin{center}
\leavevmode
\psfig{file=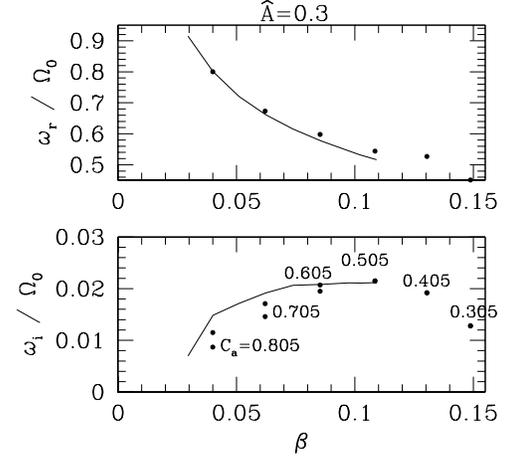,width=2.8in,angle=0}
\end{center}
\vspace*{-8mm}
\caption{The same as Fig. 2, but for $\hat A=0.3$.
For $C_a \geq 0.605$, 
upper and lower bounds of $\omega_i$ for 
results in numerical simulations are shown. 
}
\end{figure}
\begin{figure}
\vspace*{-6mm}
\begin{center}
\leavevmode
\psfig{file=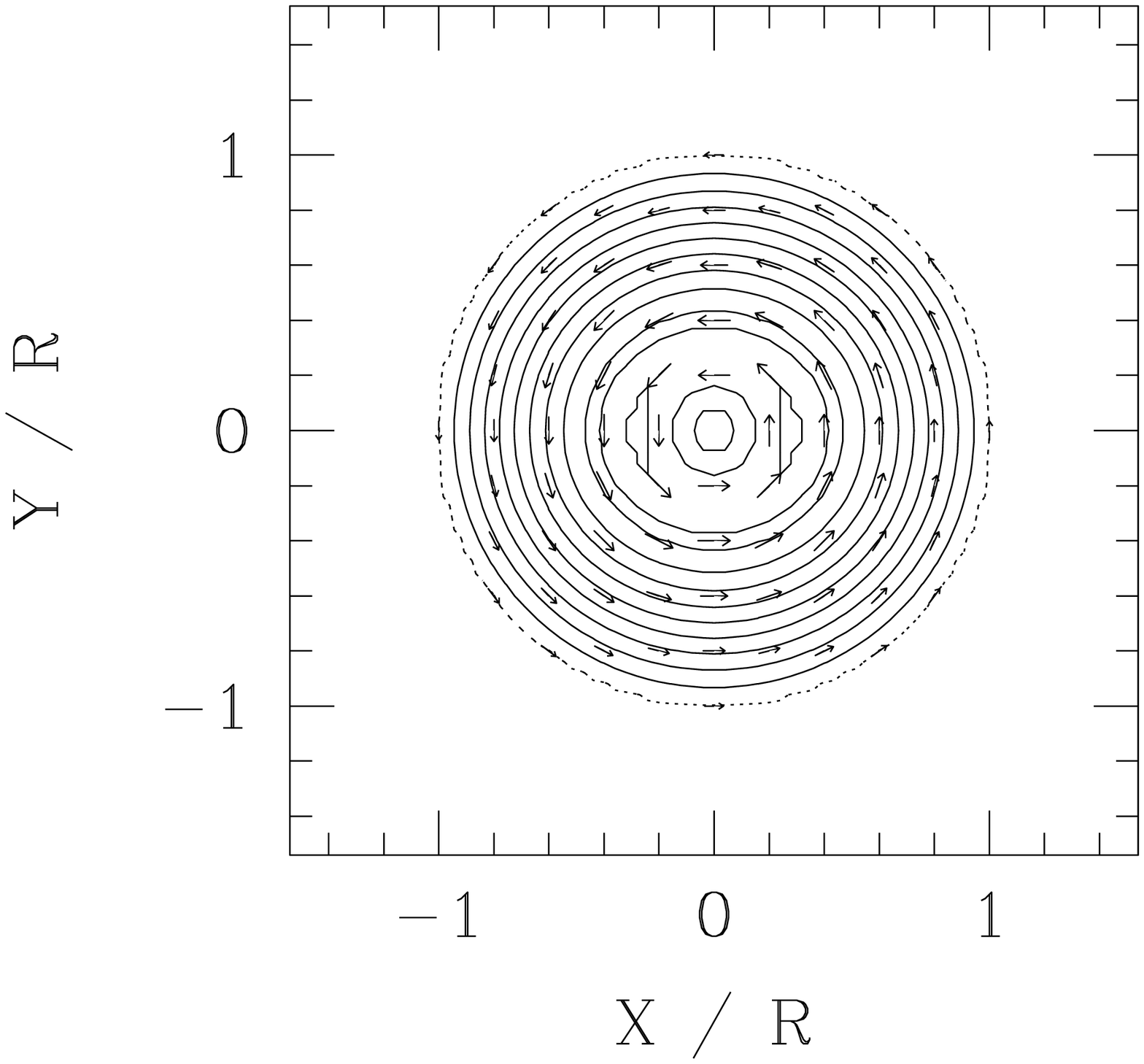,width=2.8in,angle=0}
\leavevmode
\psfig{file=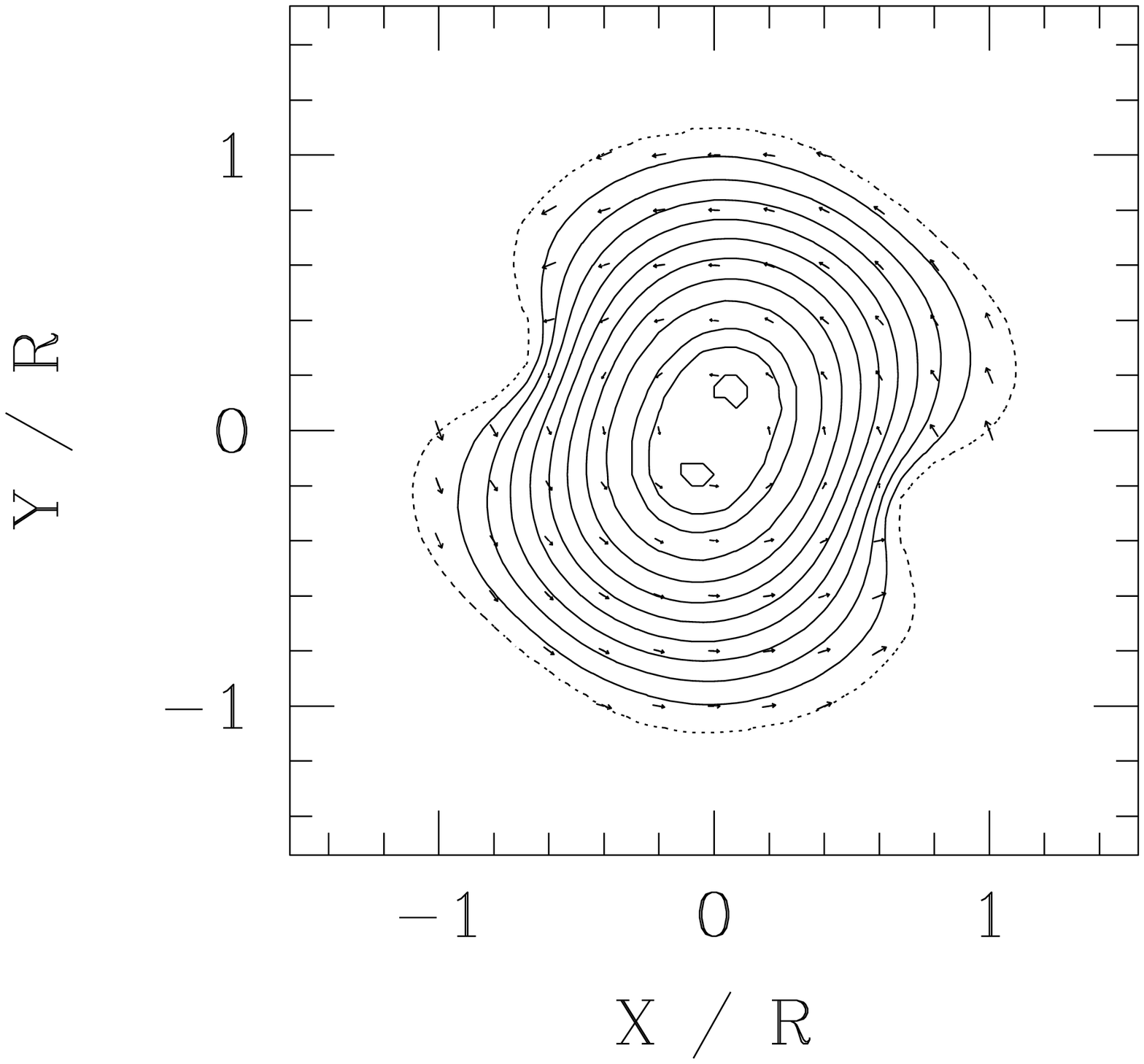,width=2.8in,angle=0}
\end{center}
\vspace*{-4mm}
\caption{Snapshots of density contour lines and velocity vectors in the
equatorial plane at $t=0$ (left) and
$\Omega_0 t \approx 390$ (right) for $\hat A=0.3$ and $C_a=0.605$.
The contour lines are drawn for $\rho/\rho_{\rm max}
=j/10~(j=1 \sim 9)$ and 0.95 (solid curves) and
for $\rho/\rho_{\rm max}=0.01$
(dotted curve), where $\rho_{\rm max}$ denotes the maximum density.
}
\end{figure}

In Fig. 5, we display the contour line and velocity vectors 
for $C_a=0.605$ at selected time slices as an example. 
We find that the dynamical instability does not excite large
spiral arms and fragmentations. Rather, it saturates 
in a weakly nonlinear stage at which $|\eta| \sim 0.2$, 
to form a quasistationary ellipsoid. We note that this feature is
independent of $C_a$ as long as $\hat A=0.3$.

Since the rotating stars of 
$C_a \alt 0.85$ 
change to nonaxisymmetric objects, 
they can be sources of gravitational waves. In Fig. 6,
we show the gravitational waveforms along the $z-$axis 
($h_{+,\times}$) and energy luminosity ($\dot E$) 
as a function of time for $C_a=0.605$. We note that the 
results for other models are qualitatively the same as that in Fig. 6.
Here, gravitational waves are calculated in the quadrupole formula 
\cite{MTW}, and we define
\beqn
h_+ \equiv {\ddot I_{xx} - \ddot I_{yy} \over r},~
h_{\times} \equiv {2\ddot I_{xy} \over r},~
{\rm and}~\dot E \equiv {1 \over 5} \bI_{ij}^{(3)}\bI_{ij}^{(3)}
\eeqn
where $\ddot I_{ij}=d^2 I_{ij}/dt^2$, $\bI_{ij}=I_{ij}-\delta_{ij} I_{kk}/3$,
$\bI_{ij}^{(3)}=d^3 \bI_{ij}/dt^3$, and
$r$ is the distance from a source to a detector.
Here, we use the units $c=G=1$. 
Since the outcome of the dynamical instability is 
a quasistationary ellipsoid, the amplitude and luminosity of gravitational
waves settle down to a nearly constant value as
$rh_{+,\times} \sim 0.2(M^2/R)$ and $\dot E \sim 0.01 (M/R)^5$.

\begin{figure}
\vspace*{-6mm}
\begin{center}
\leavevmode
\psfig{file=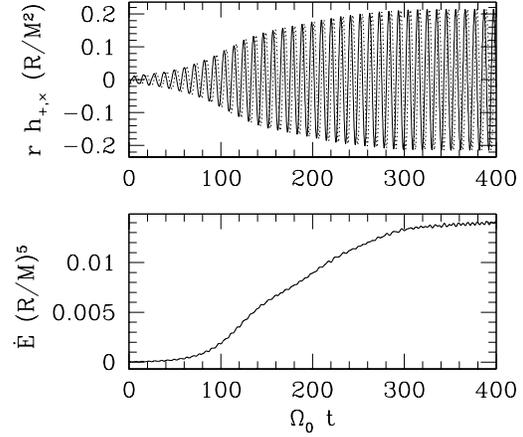,width=2.8in,angle=0}
\end{center}
\vspace*{-8mm}
\caption{Gravitational waves for $\hat A=0.3$ and $C_a=0.605$. 
Upper figure: $h_{+}$ (solid curve) and $h_{\times}$
(dotted curve) in units of $M^2/R$ as a function of $\Omega_0 t$.
Lower figure: $\dot E$ in units of $(M/R)^5$. 
}
\end{figure}

\section{DISCUSSIONS}

Using the results shown in Fig. 6, 
we can estimate the expected amplitude of gravitational waves
from the nonaxisymmetric outcome formed after the dynamical instability
saturates. Here, we pay particular attention to 
protoneutron stars likely formed soon after the supernovae of mass
$\sim 1.4M_{\odot}$ and radius several 10 km. 
For the maximum amplitude of gravitational waves 
$h \sim 0.2M^2/Rr$, and the energy luminosity 
$\dot E \sim 0.01(M/R)^5$, 
the emission timescale of gravitational waves can be estimated as
\beq
\tau \sim {T \over \dot E} \sim 100 \beta' \biggl({R \over M}\biggr)^4 M, 
\eeq
where we set $T = \beta' M^2/R$ and $0.04 \alt \beta' \alt 0.4$
for $\hat A=0.3$ and $0.3 \leq C_a \leq 0.8$. 
The characteristic frequency of gravitational waves is denoted as 
\beqn
f \equiv {\omega_r \over 2\pi} 
\approx 800 {\rm Hz}~ \biggl({\xi \over 2}\biggr)
\biggl({15M \over R}\biggr)^{3/2}
\biggl({M \over 1.4M_{\odot}}\biggr)^{-1}, 
\eeqn
where $\xi\equiv\omega_r\sqrt{R^3/M} \sim 2$ (see Figs. 4 and 7). Since 
the cycles of gravitational wave-train $N$ are estimated as
$N \equiv f\tau$, 
the effective amplitude of gravitational waves
$h_{\rm eff} = \sqrt{N}h $
is 
\beqn
h_{\rm eff}
\approx
5 \times 10^{-22} \biggl({\xi \over 2}\biggr)^{1/2}
\biggl({\beta' \over 0.1}\biggr)^{1/2}
\biggl({R \over 15M}\biggr)^{1/4} \nonumber \\
\hskip 2cm \times \biggl({M \over 1.4M_{\odot}}\biggr)
\biggl({100{\rm Mpc} \over r}\biggr) 
\eeqn
\cite{Kip,LS,LL1,LL2}.
Thus, gravitational waves from protoneutron stars of
a high degree of differential rotation and of mass
$\sim 1.4 M_{\odot}$ and radius several $10$ km in the
distance of $\sim 100$ Mpc can be a source for 
laser interferometric detectors such as LIGO \cite{Kip2}.
We emphasize that $f$ and $h_{\rm eff}$ depend weakly on 
$\beta$ (or $\beta'$) for $\hat A=0.3$. This implies that even if 
the star is not rapidly rotating, the dynamical instability 
can set in and as a result, differentially rotating 
stars can emit gravitational waves of a large amplitude.

\begin{figure}
\vspace*{-5mm}
\begin{center}
\leavevmode
\psfig{file=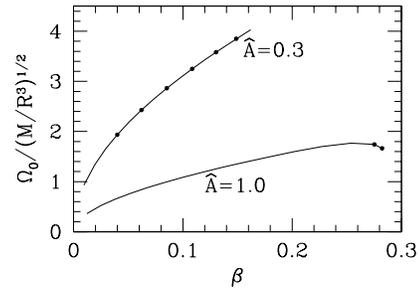,width=2.5in,angle=0}
\end{center}
\vspace*{-20mm}
\caption{$\Omega_0/\sqrt{M/R^3}$ as a function of $\beta$. 
The solid circles denote the dynamically unstable stars we found. 
}
\end{figure}

To summarize, we have studied dynamical bar-mode instability of 
differentially rotating stars focusing on the 
f-mode. We have found that rotating stars of a high degree of
differential rotation are
dynamically unstable against nonaxisymmetric deformation
even for $\beta \ll 0.27$. 


It is worthy to note that the real parts of the eigen frequencies do not 
vanish but approach to finite values as the value of $\beta$ decreases, 
i.e. in the spherical limit. This behavior 
is totally different from those of the r-modes (see e.g. Karino et al. 2001),
several self-gravity induced instability modes of slender tori or annuli
such as I-modes and J-modes \cite{GN,AT}, or shear instability modes
(or P-modes) such as Papaloizou-Pringle instability for toroids \cite{PP} 
and for spheroids \cite{Lu}.
Therefore, we have identified the unstable modes we find in this paper
as the f-mode. 

The physical mechanism for the onset of dynamical instabilities
found in this Letter may be explained in the following manner:  
For a small value of $\hat A$, non-spherical deformation of a stellar 
configuration is significant around the rotational axis although 
the distant part from the rotational axis is almost spherical.
Thus, with decrease of the value of $C_a$, most rotational energies are 
confined to the region near the rotational axis.
Increase of the rotational energy 
can be much larger than those of the gravitational and internal energies, 
because the overall shape cannot be much different from that of a sphere. 
As a result, the total energy becomes large with small values of $C_a$.
However, if the rotational energy exceeds a certain amount
in the region near the rotational axis, there may exist other equilibrium 
configurations with lower total energies, as
in the case of bifurcation of the Jacobi ellipsoidal sequence
from the Maclaurin sequence \cite{CH69}. In fact, the total energy of 
the Jacobi ellipsoid is lower than the Maclaurin spheroid 
of the same mass and the angular momentum, if
the rotational energy exceeds a certain criterion. 

Once the instability sets in, the axisymmetric star begins to change
its shape to a nonaxisymmetric configuration. In such case, 
the angular velocity in the region near the rotational axis, 
$\sim \Omega_0$, is very large (larger than $\sqrt{M/R^3}$). 
It implies that, despite of the large rotational energy, the angular 
momentum cannot be very large because $ T \sim \int (x^2 + y^2) \times 
\Omega_0^2 dm $ and $J \sim \int (x^2 + y^2) \times \Omega_0 dm$,
where $J$ and $dm$ are the angular momentum and the mass element, respectively.
Therefore, even if the nonaxisymmetric mode grows due to the
dynamical instability, the
configuration cannot be highly elongated or form spiral-arm structures; i.e., 
there exists a saturation state of a small nonaxisymmetric deformation.  
For the bar-mode instability of the Maclaurin spheroids, the rotational 
energy is "confined" to the outer part of the configuration by making the
shape very flat and so the angular momentum can be also large, even though
the angular velocity is not extremely large ($<\sqrt{M/R^3}$).
Therefore, once a bar-mode dynamical 
instability sets in, configurations can be considerably 
different from the original spheroidal shapes.


It is possible that other non-fundamental modes have larger $\omega_i$ 
than the f-mode has. However, the present study shows that 
there exists at least one mode which induces the nonaxisymmetric 
deformation. Such rotating stars subsequently form nonaxisymmetric 
structures and hence can be a source of laser interferometric
gravitational wave detectors. 

\section*{Acknowledgments}

We thank Lee Lindblom for discussion and comments.
We are grateful to Luciano Rezzolla for his careful reading and comments.
Numerical simulations were performed on FACOM VPP5000 in the 
data processing centre of National Astronomical Observatory of Japan. 
This work was in part supported by a Japanese 
Monbu-Kagaku-sho Grant (No. 12640255 and 13740143). 
SK is supported by JSPS Research Fellowship for Young Scientists.

\label{lastpage}

\end{document}